\documentclass[showpacs,superscriptaddress,prl,twocolumn]{revtex4}
\usepackage{amsmath,amssymb,amsfonts,amsthm}

\newtheorem{theorem}{Theorem}

\newtheorem{corollary}{Corollary}

\newcommand{\dv}{\, d\mu}

\newcommand{\Rt}{\mathbb{R}^3}

\newcommand{\mf}{\mathcal{M}}

\newcommand{\vv}{\alpha}

\newcommand{\vY}{y}

\newcommand{\vh}{\varphi}

\newcommand{\cbn}[1]{||#1||_{{C^{1}_\beta}(\Omega)}}

\newcommand{\bs}{\mathcal{B}}

\newcommand{\bn}[1]{||#1||_{\bs}}

\newcommand{\dy}{\Omega_{\rho_0}}

\begin{document}
\title{Angular momentum-mass inequality for axisymmetric black holes}

\author{Sergio Dain}
\email[E-mail: ]{dain@aei.mpg.de}
\affiliation{Albert-Einstein-Institut, am M\"uhlenberg
    1, D-14476, Golm, Germany.} 

\date{\today}
\begin{abstract}  
  The inequality $\sqrt{J}\leq m$ is proved for vacuum, asymptotically
  flat, maximal and axisymmetric data close to extreme Kerr data.  The
  physical significance of this inequality and its relation to  the
  standard picture of the gravitational collapse are discussed.
\end{abstract}


\maketitle

\emph{Introduction.}  --- The following conjectures constitute the
essence of the current standard picture of the gravitational collapse:
i) Gravitational collapse results in a black hole (weak cosmic
censorship) ii) The spacetime settles down to a stationary final
state. If we further assume that at some finite time all the matter
fields have fallen into the black hole and hence the exterior region
is pure vacuum (for simplicity we discard electromagnetic fields in
the exterior), then the black hole uniqueness theorem implies that the
final state should be the Kerr black hole. The Kerr black hole is
uniquely characterized by its mass $m_0$ and angular momentum
$J_0$. These quantities   satisfy the following
remarkable inequality
\begin{equation}
 \label{eq:14i}
\sqrt{|J_0|}\leq m_0.
\end{equation}
From Newtonian considerations, we can interpret this inequality as
follows\cite{Wald71}: in a collapse the gravitational attraction
($\approx m_0^2/r^2$) at the horizon ($r \approx m_0 $) dominates over
the centrifugal repulsive forces ($\approx J_0^2/m_0r^3$).

If the initial conditions for a collapse violate \eqref{eq:14i} then
the extra angular momentum should be radiated away in gravitational
waves. However, in an axially symmetric spacetime the angular momentum is a
conserved quantity (the Komar integral of the Killing vector, see, for
example, \cite{Wald84}). In this case angular momentum can not
be radiated: the angular momentum $J$ of the initial conditions must
be equal to the final one $J_0$. On the other hand, the mass of the
initial conditions $m$ satisfies $m\geq m_0$ because gravitational
radiation carries  positive energy. Then, from inequality
\eqref{eq:14i} we obtain
\begin{equation}
 \label{eq:14a}
\sqrt{|J|}\leq m.
\end{equation}
More precisely, i)-ii) imply that a complete, vacuum,
axisymmetric, asymptotically flat data should satisfy inequality
\eqref{eq:14a}, where $m$ and $J$ are the mass and angular momentum of
the data. Moreover, the equality in \eqref{eq:14a} should imply that
the data are an slice of extreme Kerr. This is a similar argument to
the one used by Penrose \cite{Penrose69} to obtain the inequality
between mass and the area of the horizon on the initial data. As in
the case of Penrose inequality, a counter example of \eqref{eq:14a}
will imply that either i) or ii) is not true. Conversely a proof of
\eqref{eq:14a} gives indirect evidence of the validity of i)-ii),
since it is very hard to understand why this highly nontrivial
inequality should hold unless i)-ii) can be thought of as providing
the underlying physical reason behind it (see the discussion in
\cite{Wald99}). The main result of this letter is that \eqref{eq:14a}
is true for data close enough to extreme Kerr data.

Inequality \eqref{eq:14a} is a property of the spacetime and not only
of the data, since both quantities $m$ and $J$ are independent of the
slicing.  It is in fact a property of axisymmetric, vacuum, black
holes space-times, because a non zero $J$ (in vacuum) implies a non
trivial topology on the data and this is expected to signal the
presence of a black hole. The physical interpretation of
\eqref{eq:14a} is the following: if we have an stationary vacuum black
hole (i.e. Kerr) and add to it axisymmetric gravitational waves, then
the spacetime will still have a (non-stationary) black hole, these
waves will only increase the mass and not the angular momentum of the
spacetime because they are axially symmetric.  Since inequality
\eqref{eq:14i} is satisfied for Kerr we get \eqref{eq:14a}.

The Kerr black hole has been proved to be unique among stationary
solutions (see the review articles \cite{Carter99} \cite{Chrusciel02c}
and references therein).  There exists also linear stability studies
for Kerr \cite{Whiting89} \cite{Beyer01}. The result presented here is
the first non linear one which proves the relevance of Kerr among
non stationary solutions of Einstein equations.

\emph{The variational principle.}  --- Inequality \eqref{eq:14a}
suggests the following variational principle: \emph{The extreme Kerr
  initial data is the absolute minimum of the mass among all
  axisymmetric, vacuum, asymptotically flat and complete initial data
  with fixed angular momentum.}  With the extra assumption that the
data are maximal, this variational principle can be formulated in a
remarkable simple form \cite{Dain05c}. A maximal initial data set for
Einstein's vacuum equations consists in a Riemannian metric
$\tilde{h}_{ab}$, and a trace-free symmetric tensor field
$\tilde{K}^{ab}$ such that the vacuum constraint equations
\begin{align}
 \label{const1}
 \tilde D_b \tilde K^{ab} = 0,\\
 \label{const2}
 \tilde R -\tilde K_{ab} \tilde K^{ab}=0,
\end{align}
are satisfied; where $\tilde{D}_a$ and $\tilde R$ are the Levi-Civita
connection and the Ricci scalar associated with $\tilde{h}_{ab}$. In
these equations the indexes are moved with the metric $\tilde h_{ab}$
and its inverse $\tilde h^{ab}$.

We  assume that the initial data are axially symmetric, that is,
there exists an axial Killing vector $\eta^a$ such that
 \begin{equation}
  \label{eq:8b}
 \pounds_\eta \tilde h_{ab}=0, \quad  \pounds_\eta \tilde K_{ab}=0,
\end{equation}
where $\pounds$ denotes the Lie derivative. 
The Killing vector $\eta^a$ is assumed to be hypersurface orthogonal
on the data.
Under these conditions, the metric $\tilde h_{ab}$ can be
characterized by two functions $q,v$; we specify them using Brill's
ansatz \cite{Brill59}.   Let $\rho,z,\phi$ be cylindrical
coordinates in $\Rt$. We
write the metric in the form
\begin{equation}
  \label{eq:1b}
  \tilde h_{ab}=e^{v}h_{ab},
\end{equation}
where the conformal metric $h_{ab}$ is given by 
\begin{equation}
  \label{eq:105}
  h= e^{-2q}(d\rho^2+dz^2)+\rho^2d\phi^2.
\end{equation} 
In these coordinates we have $\eta^a= (\partial/\partial \varphi)^a$
and its  norm with respect to the metric $\tilde h_{ab}$ will be denoted by
$X$ 
\begin{equation}
  \label{eq:72}
X =\eta^a\eta^b\tilde h_{ab}=e^v\rho^2.
\end{equation}
The function $q$ is assumed to be smooth with respect to the
coordinates $\rho,z$. At the axis we impose the regularity
condition
\begin{equation}
  \label{eq:9b}
  q(\rho=0,z)=0.
\end{equation}
This condition implies that the conformal metric $h_{ab}$ is well
defined in $\Rt$.  At infinity we assume the following fall-off
\begin{equation}
  \label{eq:10b}
  q=o(r^{-1}), \quad q_{,r}=o(r^{-2}).
\end{equation}
These fall off conditions imply that the mass of the physical metric
$\tilde h_{ab}$ is contained in the function $v$.  This function is
allowed to be singular at some points at the axis, these singularities
represent the extra asymptotic ends of the data.  In the present case,
since we are going to study small deviation from the extreme Kerr
data, we only have one extra end, the corresponding singular
point of $v$ will be chosen to be at the origin. 

The relevant part of the second fundamental form $\tilde K^{ab}$ is
characterized by a potential $Y$. Define the vector $\tilde S^a$ by
\begin{equation}
  \label{eq:2t}
  \tilde S_a=\tilde K_{ab}\eta^b-X^{-1}\tilde \eta_a \tilde
  K_{bc}\eta^b\eta^c, 
\end{equation} 
where $\tilde \eta_a=\tilde h_{ab}\eta^b$.  Using equations
\eqref{const1}, \eqref{eq:8b}  and the Killing equation we obtain
\begin{equation}
  \label{eq:73}
\tilde D_{[b}K_{a]}=0, \quad K_a= \tilde \epsilon_{abc}\tilde S^b \eta^c,
\end{equation}
 where $\tilde \epsilon_{abc}$
is the volume element of $\tilde h_{ab}$.
Then there exist a scalar function $Y$ such that $K_a=\tilde D_aY$. 
This function contains the angular momentum $J$ of the data
\begin{equation}
  \label{eq:74}
  J=\frac{1}{8}\left (Y(\rho=0,-z) -Y(\rho=0, z)\right ), \quad z\neq 0.
\end{equation} 
Summarizing, for any data $(\tilde{h}_{ab},\tilde{K}^{ab})$ which
satisfy the assumptions above, we have a pair $(v,Y)$. These functions
will be our fundamental variables. For a given $(v,Y)$ we can calculate
$q$ from the constrain equation \eqref{const2} (see the discussion in
\cite{Dain05c}). Consider the functional defined in
\cite{Dain05c}
\begin{equation}
  \label{eq:5c}
 \mf(v,Y)= \frac{1}{32\pi}\int_{\mathbb{R}^3}
  \left(|\partial  v|^2  + \rho^{-4} e^{-2v} |\partial Y |^2  \right) \dv, 
\end{equation}
where $\dv=\rho \,dz d\rho d\phi$ is the volume element in $\Rt$ and
$\partial$ denotes partial derivative with respect to $\rho$ and $z$;
that is $|\partial v|^2= v^2_{,z} +v^2_{,\rho}$. Note that this
functional does not depends on $q$.
In \cite{Dain05c} it has been proved that the following bound holds
for every maximal data 
\begin{equation}
  \label{eq:75}
m\geq \mf.
\end{equation}
Equation \eqref{eq:75} allows us to formulate the variational
principle mentioned above in terms of the functional $\mf$ which
depends only on two free functions $(v,Y)$: we want to prove that
extreme Kerr is a minimum of $\mf$ among all $(v,Y)$ which satisfy the
following boundary conditions.

Let $v_0$ and $Y_0$ denote the extreme Kerr initial data. These are
explicit functions (see \cite{Dain05d}) which depend on a free
parameter $J$, the angular momentum of the data. 
As it was pointed out above, the function $v_0$ is singular at the
origin since extreme Kerr data has two asymptotic ends, nevertheless
the mass functional \eqref{eq:5c} is finite and gives the total mass
of the extreme Kerr data  $\mf(v_0,Y_0)=\sqrt{|J|}$.  
 Set
\begin{equation}
  \label{eq:1}
v=v_0+\vv , \quad Y=Y_0+\vY.  
\end{equation} 
The functions $(\vv, \vY)$ are required to have a fall off compatible
with asymptotic flatness.  For $\vY$ we need also to prescribe boundary
conditions at the axis in order to impose that the angular momentum of
$Y$ is the same as the one of $Y_0$. From equation \eqref{eq:74} we
get that $\vY$ should vanishes at the axis.  
 To simplify the analysis we will further assume that $\vY$ vanishes in a
whole neighborhood of the axis. Note that for $\vv$ no extra boundary
conditions are imposed, we just require that it is a regular function
in $\Rt$.  These considerations are made precise in the
following definition of the Banach space $\bs$.  
 Let $\Omega$ be a (unbounded)
domain in $\Rt$. We introduce the  weighted spaces of $C^1$
functions in $\Omega$
\begin{equation}
  \label{eq:3}
\cbn{f} = \sup_{x\in \Omega} \left\{ \sigma^{-\beta} |f|
  +\sigma^{-\beta+1}|\partial f| \right\}, 
\end{equation}
with $\beta < -1/2$ and $\sigma=\sqrt{r^2+1}$, $r=\sqrt{\rho^2+z^2}$. 
Let $\rho_0>0$ be a constant and $K_{\rho_0}$ be the cylinder
$\rho\geq \rho_0$ in $\Rt$. We define the domain $\dy$ by
$\dy=\Rt\setminus K_{\rho_0}$.  The perturbation $y$ is assumed to
vanish in $K_{\rho_0}$.  The Banach space $\bs$ is defined by
\begin{equation}
  \label{eq:34}
\bn{\vh}\equiv ||\vv||_{C^1_{\beta}(\Rt)}+
||\vY||_{C^1_{\beta}(\dy)}.
\end{equation}

We consider $\mf$ as a functional on $\mf: \bs \rightarrow \mathbb{R}$.
To simplify the notation we will write $\vh\equiv (\vv,\vY)$ and $u_0\equiv (v_0, Y_0)$.
Our main result is given by the following theorem proved in \cite{Dain05d}. 
\begin{theorem}
 \label{t1}
The functional $\mf: \bs \to \mathbb{R}$ defined by
  \eqref{eq:5c} has a strict local minimum at $u_0$.
  That is, there exist $\epsilon>0$ such that
\begin{equation}
  \label{eq:5}
\mf(u_0+\vh)> \mf(u_0),
\end{equation}
for all $\vh\in \bs$ with $\bn{\vh}<\epsilon$ and $ \vh\neq 0$.
\end{theorem}
Using, essentially,  inequality \eqref{eq:75}, from this theorem we deduce the
following corollary.  
\begin{corollary} 
  Let $(\tilde{h}_{ab},\tilde{K}^{ab})$ be a maximal, axisymmetric,
  vacuum, initial data with mass $m$ and angular momentum $J$, such
  that the metric satisfies \eqref{eq:1b}--\eqref{eq:105} and
  \eqref{eq:9b}--\eqref{eq:10b}. Define $\vh$ as above. Then, there
  exists $\epsilon>0$ such that for $\bn{\vh}<\epsilon$ the
  inequality \eqref{eq:14a} holds. Moreover, $m=\sqrt{J}$ in this
  neighborhood if and only if the data are the extreme Kerr data.
\end{corollary}
The main ideas in the proof of theorem \ref{t1} are the following. 
 Consider the real-valued function
\begin{equation}
  \label{eq:19}
i_{\vh}(t)= \mf(u_0+t \vh).
\end{equation}
The  first variation of $\mf$ is given by $i'_{\vh}$, where prime denotes
derivate with respect to $t$. 
In \cite{Dain05c} it has been  proved that the extreme Kerr initial data is
a critical point of $\mf$, that is we have
\begin{equation}
  \label{eq:39}
i'_{\vh}(0)=0, \quad \text{ for all }\vh\in \bs. 
\end{equation}
If we compute the  second variation $ i''_\vh(t)$ it is not obvious
that it is positive at the critical point $t=0$. However,
using a remarkable identity found by Carter \cite{Carter71}, and the
equivalence (up to boundary terms)  between the functional $\mf$ and
Carter's Lagrangian (see \cite{Dain05c}) it is possible to prove (see
\cite{Dain05d})  that  
\begin{equation}
  \label{eq:27}
i''_{\vh}(0) \geq 0,  \quad \text{ for all }\vh\in \bs.
\end{equation}
Equation \eqref{eq:27} can be taken as an interpretation of Carter's
identity. This equation is a crucial necessary condition to guarantee
that $u_0$ is a local minimum, however it is not sufficient. In order
to provide a sufficient condition we need to prove that $i''_{\vh}(0)$
is coercive with respect to some appropriate norm. This last step
was done in \cite{Dain05d} (see Lemma 3.1 in this reference).
\begin{acknowledgments}
  This work has been supported by the Sonderforschungsbereich SFB/TR 7
  of the Deutsche Forschungsgemeinschaft.
\end{acknowledgments}


\end{document}